\documentclass[twocolumn]{aastex62}

\usepackage{color}

\received{March 17, 2018}
\accepted{\today}
\submitjournal{AJ}

\begin{document}

\title{Spitzer Light Curves of the Young, Planetary-Mass TW Hya Members 2MASS J11193254$-$1137466AB and WISEA J114724.10$-$204021.3}

\correspondingauthor{Adam C. Schneider}
\email{aschneid10@gmail.com}

\author{Adam C. Schneider}
\affil{School of Earth and Space Exploration, Arizona State University, Tempe, AZ, 85282, USA}

\author{Kevin K. Hardegree-Ullman}
\affiliation{Department of Physics and Astronomy, University of Toledo, 2801 W. Bancroft Street, Toledo, OH 43606, USA}

\author{Michael C. Cushing}
\affiliation{Department of Physics and Astronomy, University of Toledo, 2801 W. Bancroft Street, Toledo, OH 43606, USA}

\author{J. Davy Kirkpatrick}
\affiliation{IPAC, Mail Code 100-22, Caltech, 1200 E. California Boulevard, Pasadena, CA 91125, USA}

\author{Evgenya L. Shkolnik}
\affil{School of Earth and Space Exploration, Arizona State University, Tempe, AZ, 85282, USA}

\begin{abstract}

We present {\it Spitzer Space Telescope} time-series photometry at 3.6 and 4.5 $\mu$m of 2MASS J11193254$-$1137466AB and WISEA J114724.10$-$204021.3, two planetary-mass, late-type ($\sim$L7) brown dwarf members of the $\sim$10 Myr old TW Hya Association.  These observations were taken in order to investigate whether or not a tentative trend of increasing variability amplitude with decreasing surface gravity seen for L3-L5.5 dwarfs extends to later-L spectral types and to explore the angular momentum evolution of low-mass objects.  We examine each light curve for variability and find a rotation period of 19.39$^{+0.33}_{-0.28}$ hours and semi-amplitudes of 0.798$^{+0.081}_{-0.083}$\% at 3.6 $\mu$m and 1.108$^{+0.093}_{-0.094}$\% at 4.5 $\mu$m for WISEA J114724.10$-$204021.3.  For 2MASS J11193254$-$1137466AB, we find a single period of 3.02$^{+0.04}_{-0.03}$ hours with semi-amplitudes of 0.230$^{+0.036}_{-0.035}$\% at 3.6 $\mu$m and 0.453 $\pm$ 0.037\% at 4.5 $\mu$m, which we find is possibly due to the rotation of one component of the binary.  Combining our results with 12 other late-type L dwarfs observed with {\it Spitzer} from the literature, we find no significant differences between the 3.6 $\mu$m amplitudes of low surface gravity and field gravity late-type L brown dwarfs at {\it Spitzer} wavelengths, and find tentative evidence (75\% confidence) of higher amplitude variability at 4.5 $\mu$m for young, late-type Ls.  We also find a median rotation period of young brown dwarfs (10--300 Myr) of $\sim$10 hr, more than twice the value of the median rotation period of field age brown dwarfs ($\sim$4 hr), a clear signature of brown dwarf rotational evolution.  

\end{abstract}


\section{Introduction} \label{sec:intro}

Several large-scale surveys have found that infrared variability is common in L, T, and Y-type brown dwarfs as evinced by ground-based (\citealt{rad14}, \citealt{wil14}) and space-based {\it Spitzer} and {\it HST} surveys (\citealt{bue14}, \citealt{met15}, \citealt{cush16}).  Such variability is typically attributed to the rotational modulation of inhomogeneous cloud cover.  In the standard paradigm, the L to T transition is believed to arise from a rapid loss of the liquid iron and silicate clouds over a narrow range in effective temperature by some unknown mechanism.  This loss could produce patchy surface coverage, which would result in photometric and spectroscopic variability (\citealt{ack01}, \citealt{burg02}).  \cite{rad14} found that while variability can occur for a wide range of spectral types, high-amplitude variation ($>$2\%) is preferentially found at the L/T transition, evidence that supports this prediction.  An alternative explanation, presented in \cite{trem16}, shows that a temperature gradient reduction caused by fingering convection can reproduce the near-infrared colors of brown dwarfs across the L-T transition without the need for clouds, though \cite{lec18} argue that this mechanism cannot account for features across the L-T transition.  Analyses of brown dwarf light curves have allowed for sophisticated modeling of brown dwarf surface features (e.g., \citealt{kar15}) and the identification of zonal bands with varying wind speeds \citep{apai17}. 

\cite{met15} used {\it Spitzer} to investigate the cloud properties of a sample of 44 brown dwarfs with spectral types between L3 and T8 through a careful analysis of their [3.6] and [4.5] light curves.  One intriguing finding from that study was the tendency of low surface gravity brown dwarfs with spectral types between L3 and L5.5 to have higher amplitude variability compared to counterparts with normal surfaces gravity (i.e., field ages) in the same spectral type bin.  The three objects with the highest [3.6] amplitudes in this spectral type range all showed signatures of low surface gravity.  Even when considering the variability amplitude upper limits of low surface gravity brown dwarfs for which no variability was detected, \cite{met15} found that the correlation between low surface gravity and enhanced variability amplitudes was significant at the 92\% level, possibly indicating a link between low surface gravity and cloud structure/distribution.  Whether or not this trend extends to other spectral types is yet unknown.   

2MASS J11193254$-$1137466AB (hereafter 2MASS 1119$-$1137AB) was found in a targeted search for L and T dwarfs with unusually red colors, which are often a sign of youth for brown dwarfs (see e.g., \citealt{kirk08}, \citealt{fah13}), using SDSS, 2MASS, and {\it WISE} \citep{kell15}. This object was subsequently found to be an approximately equal magnitude binary with a separation of 0\farcs14 (3.6 $\pm$ 0.9 AU) \citep{best17}.   WISEA J114724.10$-$204021.3 (hereafter WISEA 1147$-$2040) was found as part of a larger program focused on finding young, late-type L dwarfs based on their 2MASS and AllWISE colors (\citealt{schneid16}, \citealt{schneid17}).   Both of these objects were found to have spectral types of L7, spectra with clear signs of low surface gravity (i.e., young ages), and kinematic properties consistent with membership in the TW Hya association (\citealt{kell15}, \citealt{kell16}, \citealt{schneid16}, \citealt{gagne17}). Membership is further supported by their sky positions relative to other TW Hya association members \citep{schneid16}.  The young (10 $\pm$ 3 Myr -- \citealt{bell15}) TW Hya association is one of the nearest regions of recent star formation.  Its proximity ($\sim$30-80 pc) and young age make it an excellent testbed for studying early phases of stellar and substellar evolution.  Thus 2MASS 1119$-$1137AB and WISEA 1147$-$2040 provide vital anchor points for low-mass evolutionary models and unique testbeds for investigating the atmospheres of planetary-mass objects.      

\cite{fah16} estimated the mass of WISEA 1147$-$2040 to be $\sim$6 $M_{\rm Jup}$, while \cite{best17} finds the masses of each component of 2MASS 1119$-$1137AB to be $\sim$4 $M_{\rm Jup}$.  These estimates make WISEA 1147$-$2040 and 2MASS 1119$-$1137AB the lowest mass free floating confirmed members of the TW Hya association and two of the lowest mass brown dwarfs in the Solar neighborhood.  Only the planetary mass companion 2M1207b ($\sim$5 $M_{\rm Jup}$; \citealt{cha04}, \citealt{cha05}) and the exoplanets 51 Eri b ($\sim$2 $M_{\rm Jup}$; \citealt{mac15}) and HR 8799 b ($\sim$5 $M_{\rm Jup}$; \citealt{mar08}, \citealt{mar10}), and possibly the extremely cold ($\sim$250 K), nearby ($\sim$2 pc) brown dwarf WISE 0855$-$0714 (1.5--8 $M_{\rm Jup}$; \citealt{legg17}) have been imaged directly and have comparable masses.  As such, 2MASS 1119$-$1137AB and WISEA 1147$-$2040 provide exceptional laboratories for investigating the chemistry and cloud structure in a mass and surface gravity regime not yet probed.  We have monitored 2MASS 1119$-$1137AB and WISEA 1147$-$2040 with the {\it Spitzer Space Telescope} to measure variability and to attempt to determine whether or not the trend of large amplitude variability with low surface gravity extends to later-L spectral types. 

\section{Observations of 2MASS 1119$-$1137AB and WISEA 1147$-$2040}

We used the Infrared Array Camera (IRAC; \citealt{faz04}) aboard the {\it Spitzer Space Telescope} to monitor 2MASS 1119$-$1137AB and WISEA 1147$-$2040.  2MASS 1119$-$1137AB was observed on 2017 April 24 and WISEA 1147$-$2040 was observed on 2017 April 17 (PID: 13018).  Both targets were observed for a total of 20 continuous hours; 10 hours with the 3.6 $\mu$m filter and 10 with the 4.5 $\mu$m filter (hereafter [3.6] and [4.5]) with 12 second exposures.  Following the outlined procedures for obtaining high-precision photometry from the Spitzer Science Center\footnote{https://irachpp.spitzer.caltech.edu}, science exposures were preceded by a 30 minute dither sequence to account for initial slew settling and were also followed by a 10 minute dither sequence.  Science exposures were taken with the target located on the ``sweet spot'' of the detector, a region used to minimize correlated noise.  Limiting our AOR lengths to 10 hours ensures that any {\it Spitzer} pointing system drift, which is typically $\sim$0\farcs35/day \citep{grill12, grill14}, does not shift our targets away from the well-characterized detector sweet spot.    

We use the \texttt{photutils} package \citep{Bradley2016} for centroiding and aperture photometry. Each image is cropped to a $32 \times 32$ pixel region around the sweet spot and a 2D Gaussian is fit to find the centroid. We then extract photometry for each pixel in the $5 \times 5$ pixel region around the centroid. Background levels were found using the method described by \cite{Knutson2011}. We adapt the pixel-level decorrelation (PLD) method used by \cite{Benneke2017} and originally developed by \cite{Deming2015} to account for intrapixel sensitivity variations in the Spitzer/IRAC photometry. The instrument sensitivity $S(t_i)$ can be modeled with 25 time-independent pixel weights $w_k$ using

\begin{equation}
    S(t_i)=\frac{\sum\limits^{25}_{k=1} w_k D_k(t_i)}{\sum\limits^{25}_{k=1} D_k(t_i)}+m \cdot t_i,
\end{equation}

\noindent where $D_k(t_i)$ are the number of electrons in each pixel $k$ at time $t_i$, and $m$ is a linear slope. We then fit the raw {\it Spitzer} photometry using the log-likelihood function

\begin{equation}
    \ln \mathcal{L} = -\frac{1}{2}\sum\limits^N_{i=1}\bigg[\bigg(\frac{\sum^{n}_{k=1} D_k(t_i) - S(t_i)}{\sigma}\bigg)^2+\ln(2\pi\sigma^2)\bigg],
\end{equation}

\noindent where $\sigma$ is the photometric scatter fit simultaneously with the instrument systematic model. This differs from \citet{Benneke2017} Equation 2 in that we do not include an astrophysical model. Since the shape of the astrophysical signal is not known beforehand, we do not want to introduce a spurious signal. Therefore, we fit the astrophysical model separately as described in the following section. We use the {\tt\string emcee} package \citep{fm13}, which applies the affine-invariant ensemble sampler of \cite{good10} to implement a Markov Chain Monte Carlo (MCMC) approach to find posterior distributions using the likelihood function above. For each fit we use 100 walkers with 30,000 steps, where the first 10,000 steps are burn-in. We apply the pixel weights to the input photometry to yield a corrected flux used in the analysis below.

\section{Analysis}
We analyzed the [3.6] and [4.5] observations for 2MASS 1119$-$1137AB and WISEA 1147$-$2040 with a probabilistic model defined as:

\begin{equation}
D_i = C + A\sin\left(\frac{2\pi}{P}t_i + \phi\right) + \epsilon,
\end{equation}

\noindent where $D_i$ is the number of electrons detected at time $t_i$, $C$ is an additive constant to account for any shift in the Y direction, $A$ is the amplitude, $P$ is the period, $\phi$ is the phase, and $\epsilon$ is the measurement error.  We again used the {\tt\string emcee} package to fit each set of data and find posterior distributions for each of the above model parameters.  We run 1000 walkers with 1000 steps for each fit, where the first 300 steps are treated as a burn-in sample.  

For 2MASS 1119$-$1137AB, we provide the priors used and all determined parameters from the resulting fits in Table 1 and show the fits to the data in Figure 1. Considering that the components of 2MASS 1119$-$1137AB are roughly equal mass and unresolved in our {\it Spitzer} images, it is impossible to determine how much each member of this binary is responsible for the observed variability.  We checked for variability in the residuals of both the [3.6] and [4.5] data accounting for the measured periods and amplitudes in Table 1, and found no additional variations.  We also attempted a two-component fit and were unable to identify a second rotation period.  \cite{best17} estimate an orbital period for 2MASS 1119$-$1137AB of 90$^{+80}_{-50}$ yr, so tidal-locking is unlikely.  It is possible that the variability we see originates from a single component as has been seen for the L7.5+T0.5 binary Luhman16AB \citep{burg14}.  However, without resolved images, it is not possible to determine the degree to which the second component has affected the rotational parameters we have measured from this pair's combined light curve.  For this reason, we exclude the rotational parameters found for 2MASS 1119$-$1137AB from the analysis presented in Section 4.

\begin{deluxetable*}{llccccccc}
\tablecaption{emcee Best Fit Model Parameters}
\tablehead{
\colhead{} & \colhead{} & \multicolumn{2}{c}{2MASS 1119$-$1137AB} & \multicolumn{3}{c}{WISEA 1147$-$2040}\\
\cline{3-4}
\cline{5-7}
\colhead{Model Parameter} & \colhead{Prior} & \colhead{[3.6]} & \colhead{[4.5]} & \colhead{[3.6]} & \colhead{[4.5]} & \colhead{[3.6]+[4.5]\tablenotemark{b}}}
\startdata
Y-Shift & $\mathcal{U}$(0.9,1.1) & 0.9999 $\pm$ 0.0003 & 0.9998 $\pm$ 0.0003 & 1.0051 $\pm$ 0.0006 & 0.9930 $\pm$ 0.0007 & 1.000 $\pm$ 0.0002 \\
Amplitude (\%) & $\mathcal{U}$(0,10) & 0.230$^{+0.036}_{-0.035}$ & 0.453 $\pm$ 0.037 & 0.798$^{+0.081}_{-0.083}$ & 1.108$^{+0.093}_{-0.094}$ & 0.853 $\pm$ 0.029 \\
Period (hours) & $\mathcal{U}$(2,25) & 3.02$^{+0.07}_{-0.06}$ & 3.02$^{+0.04}_{-0.03}$ & 19.39$^{+0.33}_{-0.28}$\tablenotemark{a} & 19.39$^{+0.33}_{-0.28}$\tablenotemark{a} & 19.39$^{+0.33}_{-0.28}$ \\
Phase (degrees) & $\mathcal{U}$(0,360) & 29$^{+16}_{-13}$ & 101 $\pm$ 8 & 182$^{+3}_{-2}$ & 359$^{+1}_{-2}$ & 180$^{+4}_{-3}$  \\
Standard deviation & $\mathcal{U}$(0,0.5) & 0.0129 $\pm$ 0.0002 & 0.0131 $\pm$ 0.0002 & 0.0133 $\pm$ 0.0002 & 0.0150 $\pm$ 0.0002 & 0.0142 $\pm$ 0.0001\\
\enddata
\tablenotetext{a}{The period of WISEA 1147$-$2040 was fixed to this value for the [3.6] and [4.5] fits.}
\tablenotetext{b}{While the purpose of the [3.6]+[4.5] fit was solely to determine the rotation period of WISEA 1147$-$2040, we include the results for the other parameters for completeness.}
\end{deluxetable*}

\begin{figure*}
\plotone{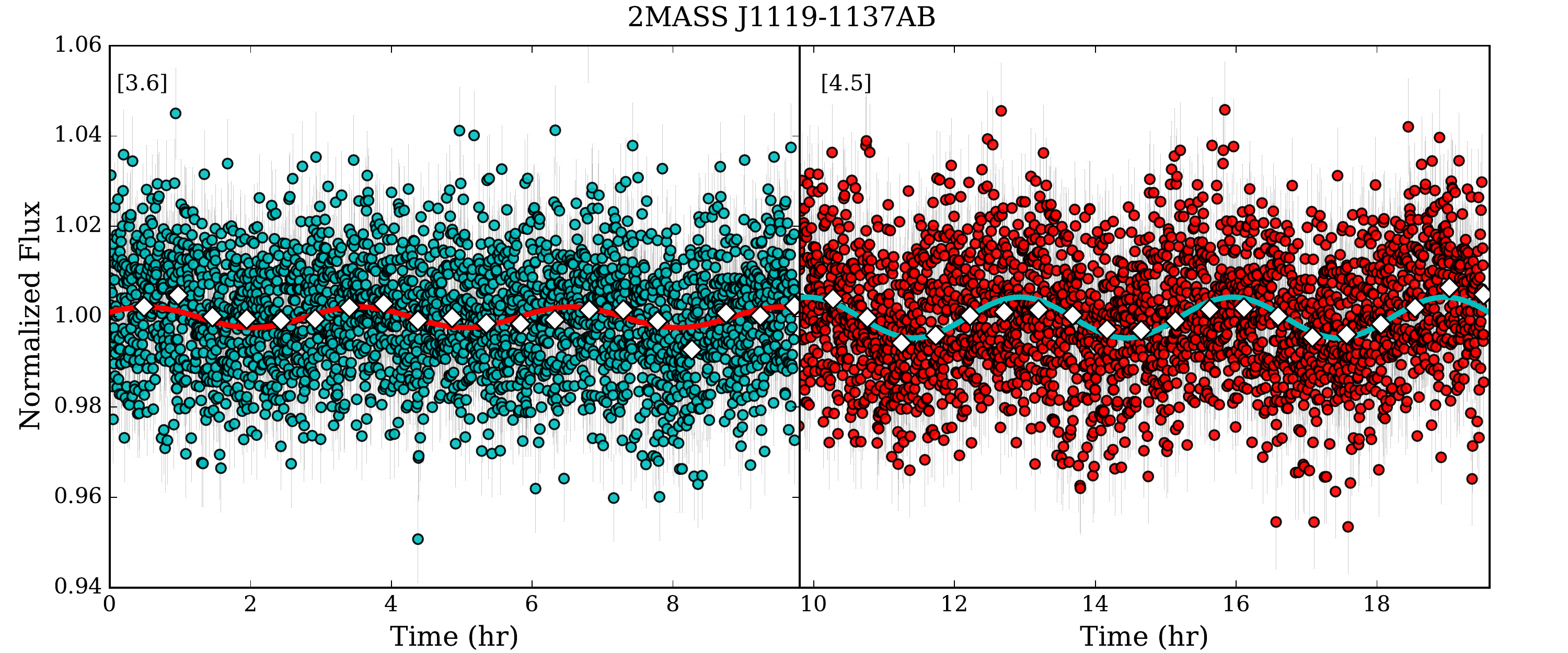}
\caption{{\it Spitzer} [3.6] (left) and [4.5] (right) light curves for 2MASS 1119$-$1137AB. The best-fit lines described in Table 1 are plotted as a red line for [3.6] and a cyan line for [4.5].  Binned median values are shown as white diamonds.}  
\end{figure*}

The measured periods for the [3.6] and [4.5] observations of WISEA 1147$-$2040 are longer than the total duration of the observations for each filter (10 hours) and the resulting posterior distributions are non-Gaussian with large uncertainties; 12.99$^{+4.18}_{-2.10}$ for [3.6] and 15.50$^{+1.67}_{-2.80}$ for [4.5].  To determine a more accurate rotation period for WISEA 1147$-$2040, we fit the [3.6] and [4.5] observation simultaneously.  Note that for all objects with regular periods in their study,  \cite{met15} found no evidence of phase shifts between [3.6] and [4.5] observations.  We scale the [3.6] and [4.5] observations by setting the median value of a set of the final [3.6] observations to equal the median value of a set of the first [4.5] observations.  We find a rotation period of 19.39$^{+0.33}_{-0.28}$ from the [3.6]+[4.5] fit. This period did not vary when the number of observations used to scale the [3.6] and [4.5] observations was between 50 and 500. To determine the variability amplitudes in the [3.6] and [4.5] wavelength regions, we then fix the period of WISEA 1147$-$2040 to the period determined from the [3.6]+[4.5] fit and rerun the {\tt\string emcee} fitting procedure outlined above.  The results of the fits to the [3.6] and [4.5] data are provided in Table 1.  The individual fits to the [3.6] and [4.5] data with a fixed period of 19.39$^{+0.33}_{-0.28}$ are shown in Figure 2, and the [3.6]+[4.5] fit is shown in Figure 3.

\begin{figure*}
\plotone{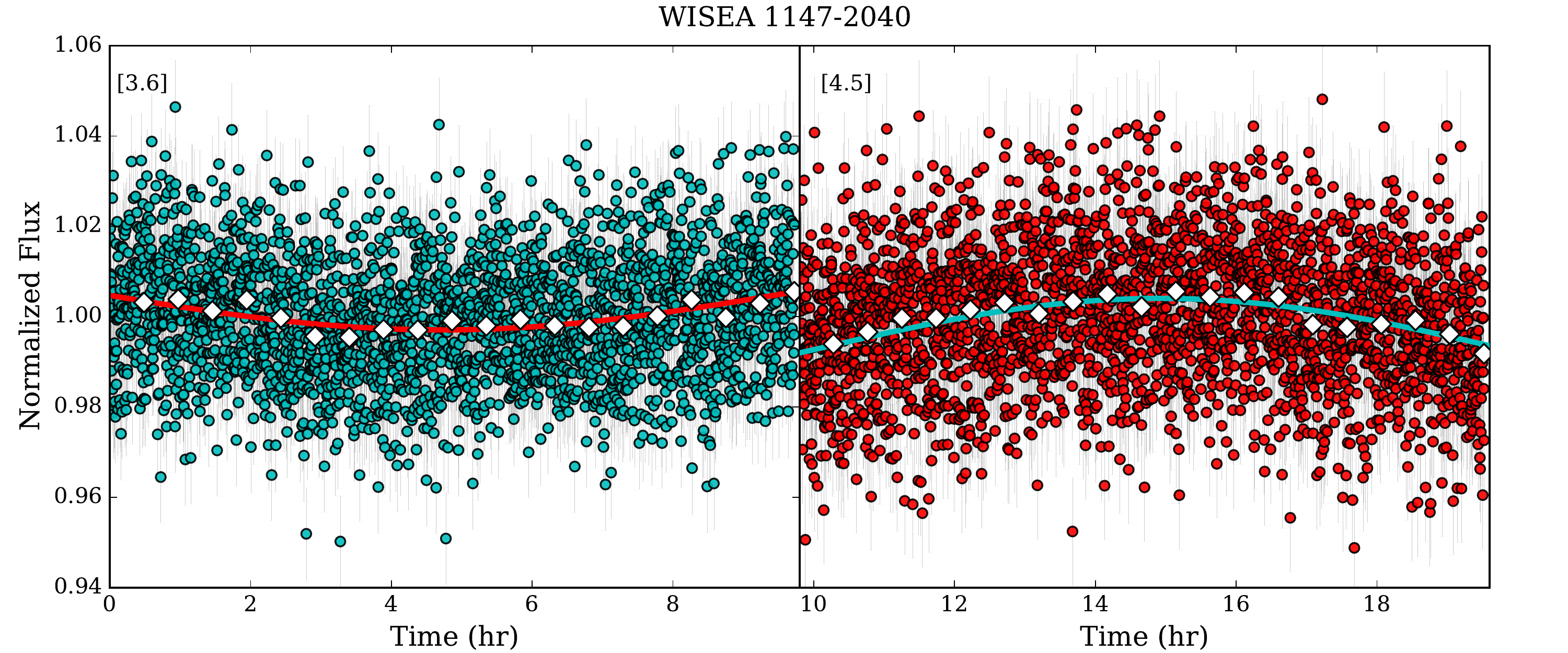}
\caption{{\it Spitzer} [3.6] (left) and [4.5] (right) light curves for WISEA 1147$-$2040 (bottom) with the period fixed at 19.39 hours. The best-fit lines described in Table 1 are plotted as red lines for [3.6] and cyan lines for [4.5].  Binned median values are shown as white diamonds.}  
\end{figure*}

\begin{figure*}
\plotone{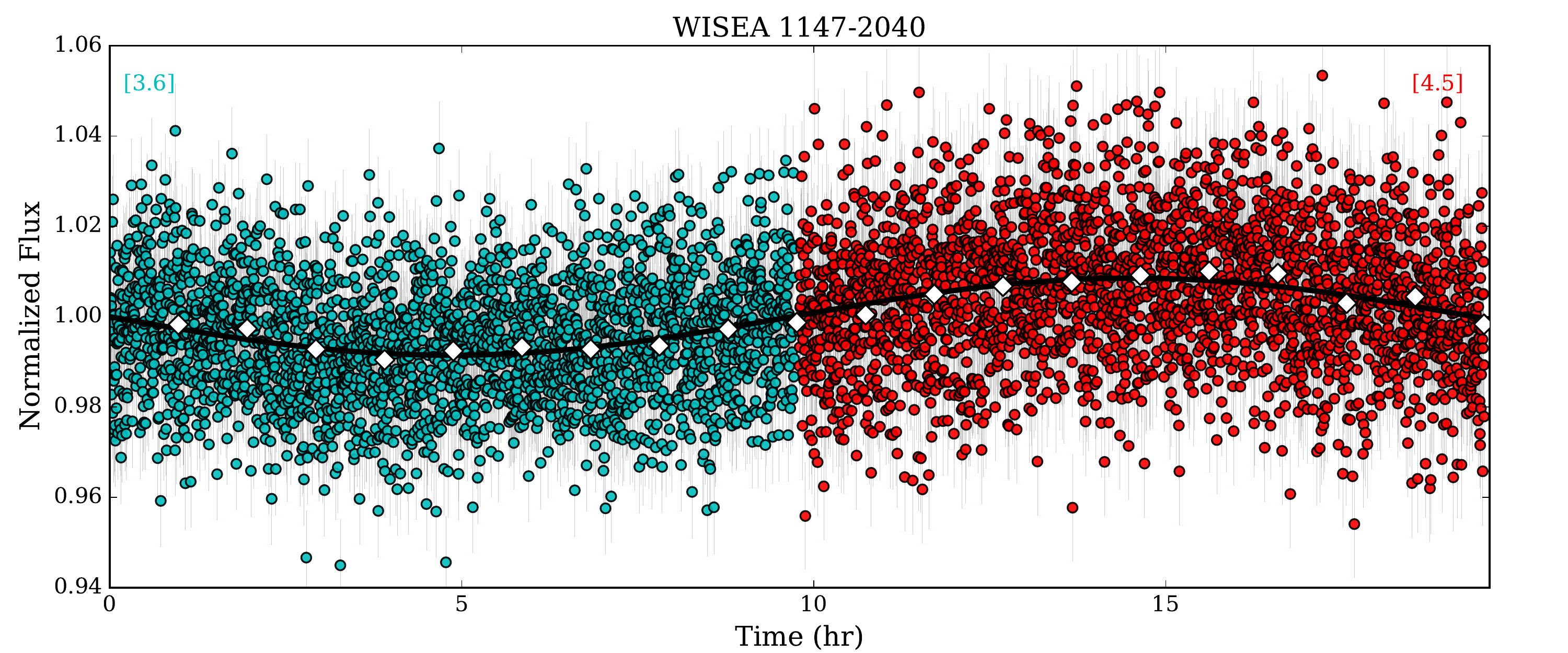}
\caption{{\it Spitzer} [3.6] and [4.5] combined light curve for WISEA 1147$-$2040. The best-fit line described in Table 1 is plotted as a black line.  Binned median values are shown as white diamonds.}  
\end{figure*}

While the {\tt\string emcee} analyses above are ideal for determining accurate rotational parameters and their uncertainties, the Lomb-Scargle periodogram (\citealt{lomb76}, \citealt{scar82}) has proven to be a powerful tool for evaluating the significance of rotation periods found in light curves as encapsulated in the false alarm probability (FAP).  We perform a period search by computing the Lomb-Scargle periodogram for the 2MASS 1119$-$1137AB [3.6] and [4.5] light curves and find peak powers at 3.02 hours for both datasets.  For the WISEA 1147$-$2040 [3.6]+[4.5] combined light curve we find a peak power at 19.23 hours, consistent with our {\tt\string emcee} fit above.  The periodogram power distributions are shown in Figure 4.  To calculate FAPs, we adopt the method of \cite{herb02}, whereby we generate 1000 artificial light curves from our data by keeping the dates the same but randomizing the measured flux values.  The tenth highest peak periodogram power from these 1000 artificial curves then defines the 1\% FAP, as only 1\% of artificial light curves would have peak powers greater than that value.   We find that no peak power of an artificial light curve approaches the peak value found for our actual light curves, which we find to be 2.5, 8.3, and 44.1 times the 1\% FAP for the 2MASS 1119$-$1137AB [3.6], 2MASS 1119$-$1137AB [4.5], and WISEA 1147$-$2040 [3.6]+[4.5] combined light curves, respectively.  Thus, we are confident that the rotation periods presented in Table 1 are significant.  

\begin{figure}
\plotone{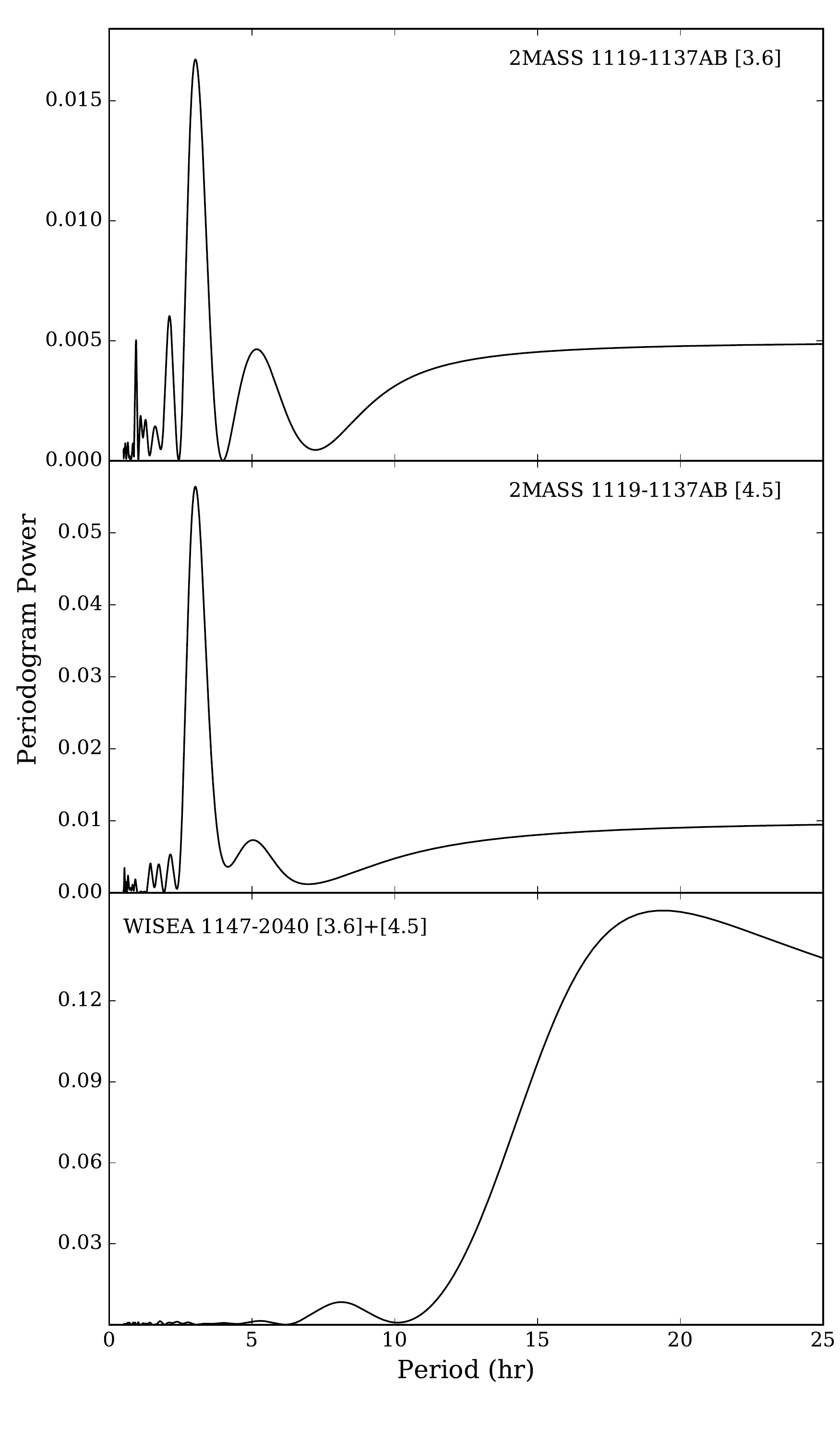}
\caption{Periodogram power distributions for the 2MASS 1119$-$1137AB [3.6] (top), 2MASS 1119$-$1137AB [4.5] (middle), and WISEA 1147$-$2040 [3.6]+[4.5] (bottom) light curves.  }  
\end{figure}

\section{Discussion}

\subsection{Variability amplitudes for L6-L9 brown dwarfs and surface gravity}

To investigate whether or not the trend of higher amplitude variability with low surface gravity seen for L3-L5.5 dwarfs in \cite{met15} continues to later L spectral types (L6-L9), we compared our light curve analysis of  WISEA 1147$-$2040 with published {\it Spitzer} light curves of other low surface gravity late-type Ls; PSO J318.5338$-$22.8603 \citep{biller18}, a member of the 23$\pm$3 Myr $\beta$ Pic association (\citealt{liu13}, \citealt{allers16}) and WISE J004701.06$+$680352.1 and 2MASS J22443167$+$2043433 \citep{vos18}, members of the $\sim$150 Myr old AB Dor moving group (\citealt{gagne14}, \citealt{giz15}).  We also use the sample of late-type Ls from \cite{met15}, which includes one low surface gravity object, 2MASSI J0103320$+$193536 (\citealt{allers13}, \citealt{mart17}), and eight late-type Ls with field gravities (ages). Table 2 lists all late-type Ls with {\it Spitzer} light curves.  Note that peak-to-peak amplitudes have been converted to semi-amplitudes.  

One object, 2MASS J21481628$+$4003593, has unusually red near-infrared colors compared to other brown dwarfs with similar spectral types.  \cite{loop08} speculate that this is due to thick dust clouds in 2MASS J21481628$+$4003593's atmosphere and find no clear evidence of youth for this object.  \cite{allers13} designate this object as a field gravity (FLD-G) source based on gravity-sensitive spectroscopic indices, though they do note that its H-band shape resembles that of low surface gravity Ls.  \cite{mart17} used medium-resolution J-band spectra to measure surface gravity sensitive indices for a large sample of M, L, and T dwarfs and designate 2MASS J21481628$+$4003593 as having an intermediate surface gravity (INT-G).  Because of the unusual cloud properties of 2MASS J21481628$+$4003593 and its uncertain surface gravity classification, we exclude it from further analysis.

\begin{deluxetable*}{lcccccccc}
\tablecaption{Late-type L (L6-L9) Rotation Properties}
\tablehead{
\colhead{Name} & \colhead{Discovery} & \colhead{Spectral\tablenotemark{a}} & \colhead{Period} & \colhead{A[3.6]} & \colhead{A[4.5]} & \colhead{Low} & \colhead{Variability} \\
\colhead{} & \colhead{Ref.} & \colhead{Type} & \colhead{(hours)} & \colhead{(\%)} & \colhead{(\%)} & \colhead{Gravity?} & \colhead{Ref.}  }
\startdata
WISE J004701.06$+$680352.1 & 1 & L7 & 16.4 $\pm$ 0.2 & 0.54 $\pm$ 0.02 & \dots & Y & 15 \\
2MASSI J0103320$+$193536 & 2 & L6 & 2.7 $\pm$ 0.1 & 0.28 $\pm$ 0.02 & 0.44 $\pm$ 0.05 & Y & 16 \\
SDSSp J010752.33$+$004156.1 & 3 & L8 & 5.0\tablenotemark{b} & 0.64 $\pm$ 0.07 & 0.5 $\pm$ 0.1 & N & 16 \\
2MASSI J0825196$+$211552 & 2 & L7.5 & 7.6\tablenotemark{b} & 0.41 $\pm$ 0.04 & 0.7 $\pm$ 0.2 & N & 16 \\
SDSS J104335.08$+$121314.1 & 4 & L9 & 3.8 $\pm$ 0.2\tablenotemark{b} & 0.77 $\pm$ 0.08 & 0.6 $\pm$ 0.1 & N & 16 \\
2MASS J11193254$-$1137466AB\tablenotemark{c} & 5 & L7 & 3.02$^{+0.04}_{-0.03}$ &  0.230$^{+0.036}_{-0.035}$ & 0.453 $\pm$ 0.037 & Y & 17 \\
WISEA J114724.10$-$204021.3 & 6 & L7 & 19.39$^{+0.33}_{-0.28}$ & 0.798$^{+0.081}_{-0.083}$ & 1.108$^{+0.093}_{-0.094}$  & Y & 17 \\
SDSS J141624.08$+$134826.7 & 7,8,9,10 & L6+T7.5 & \dots & $<$0.08 & $<$0.11 &N & 16 \\
SDSS J154508.93$+$355527.3 & 4 & L7.5 & \dots & $<$0.30 & $<$0.58 & N & 16 \\
2MASSW J1632291$+$190441 & 11 & L8 & 3.9 $\pm$ 0.2 & 0.21 $\pm$ 0.04 & 0.3 $\pm$ 0.2 & N & 16 \\
SDSS J204317.69$-$155103.4 & 4 & L9 & \dots & $<$0.36 & $<$0.37 & N & 16 \\
PSO J318.5338$-$22.8603 & 12 & L7 & 8.6 $\pm$ 0.1 & \dots & 1.7 $\pm$ 0.05 & Y & 18 \\
2MASS J21481628$+$4003593 & 13 & L6 & 19 $\pm$ 4 & 0.67 $\pm$ 0.04 & 0.52 $\pm$ 0.05 & N?\tablenotemark{d} & 16 \\
2MASS J22443167$+$2043433 & 14 & L6 & 11.0 $\pm$ 2.0 & 0.4 $\pm$ 0.1 & \dots & Y & 15 \\
\enddata
\tablenotetext{a}{Typical spectral type uncertainties are $\pm$0.5 subtypes.}
\tablenotetext{b}{Irregular variability or long period.}
\tablenotetext{c}{2MASS 1119$-$1137AB is included in this table for completeness, but not used in any of our analyses because of the unquantified effects of binarity on our determined rotational parameters.}
\tablenotetext{d}{2MASS J21481628$+$4003593 is thought to have an exceptionally cloudy atmosphere \citep{loop08}, and its gravity classification is unclear, with a FLD-G gravity classification in \cite{allers13} and an INT-G gravity classification in \cite{mart17}.}
\tablerefs{(1) \cite{giz12}, (2) \cite{kirk00}, (3) \cite{geb02}, (4) \cite{chiu06}, (5) \cite{kell15}, (6) \cite{schneid16}, (7) \cite{burn10}, (8) \cite{sch10}, (9) \cite{bowl10}, (10) \cite{schmidt10}, (11) \cite{kirk99}, (12) \cite{liu13}, (13) \cite{loop08}, (14) \cite{dahn02}, (15) \cite{vos18}, (16) \cite{met15}, (17) This work, (18) \cite{biller18}   }
\end{deluxetable*}

We note that low surface gravity (young) and field age late-type L brown dwarfs have different masses.  \cite{fil15} find a mass range for L6-L9 objects of $\sim$40-60 $M_{\rm Jup}$, while low-gravity late-type Ls have estimated masses of $\sim$5-15 $M_{\rm Jup}$ (\citealt{fil15}, \citealt{fah16}).  However, the effective temperatures ($T_{\rm eff}$) of low surface gravity and field gravity late-type Ls are found to be similar.  \cite{fil15} determined semi-empirical $T_{\rm eff}$ estimates by combining spectral energy distributions made with optical and infrared spectra and photometry with parallaxes and radius estimates from evolutionary models.  For field-age late-type Ls in their sample, they find a $T_{\rm eff}$ range of 1139-1518 K, while the four low surface gravity late-type Ls they studied have estimated $T_{\rm eff}$ values between $\sim$1200 and 1250 K.  Thus a comparison of these samples provides information about how cloud properties do or do not change for brown dwarfs with similar effective temperatures and different surface gravities. 

One additional consideration when discussion variability amplitudes is inclination angle, as brown dwarfs inclined such that we view them pole-on ($i$=0$\degr$) would not show variations due to rotation.  \cite{vos17} investigated the relationship between inclination angle and variability amplitude for a sample of 19 brown dwarfs with measured variability and found a clear trend of increasing of $J$-band variability amplitudes with larger inclination angles.  For {\it Spitzer} wavelengths, however, the differences between variability amplitudes of objects viewed close to equator-on ($i$$\approx$90$\degr$) was marginal compared to objects with inclinations as low as $\sim$20$\degr$.  That the $J$-band amplitudes are more affected by inclination than the {\it Spitzer} amplitudes is explained by the depths probed at these different wavelengths.  $J$-band observations probe deeper into brown dwarfs atmospheres and are therefore subject to an increased path-length through a brown dwarfs atmosphere at low inclination angles, while {\it Spitzer} wavelengths mostly probe the top of the photosphere.  In the following analysis, we ignore any effects due to inclination angle.   

Figure 5 shows a comparison of the [3.6] and [4.5] semi-amplitudes of the low surface gravity and field-age sample of late-type L dwarfs provided in Table 2.  To determine the probability that the semi-amplitudes of the young and field-age samples were drawn from the same parent sample, we first employ a Kaplan-Meier estimator \citep{kap58} using the \texttt{lifelines} Python package \citep{dav16}.  The Kaplan-Meier estimator constructs cumulative distribution functions for each sample accounting for censored data (i.e., upper limits).  We use a log-rank parametric test to evaluate the null hypothesis that these cumulative distributions have the same parent distribution.  We find $p$-values, which give the probability that these populations are not drawn from a single distribution, of 0.953 for the [3.6] sample and 0.241 for the [4.5] sample, where values $<$ 0.05 are typically interpreted as indicating two statistically distinct samples.  The $p$-value for the [3.6] data indicates no statistically significant difference between the two populations, which is clearly seen in Figure 5.  While the $p$-value for the [4.5] samples does not meet typical significance thresholds, it does suggest that there is a $\sim$75 percent chance that the differences between the two populations are not due to random chance.  We caution that the number of young objects used for the [4.5] comparison is small (3) and this result should be treated as preliminary until more data is available.  It is intriguing, however, that the two largest [4.5] amplitudes of the entire late-type L sample belong to WISEA 1147$-$2040 and PSO J318.5338$-$22.8603, both young objects. The small sample size limits the significance and hence an expanded sample of mid-infrared variability amplitudes for late-type L young brown dwarfs would help to further explore this result.

\begin{figure*}
\plotone{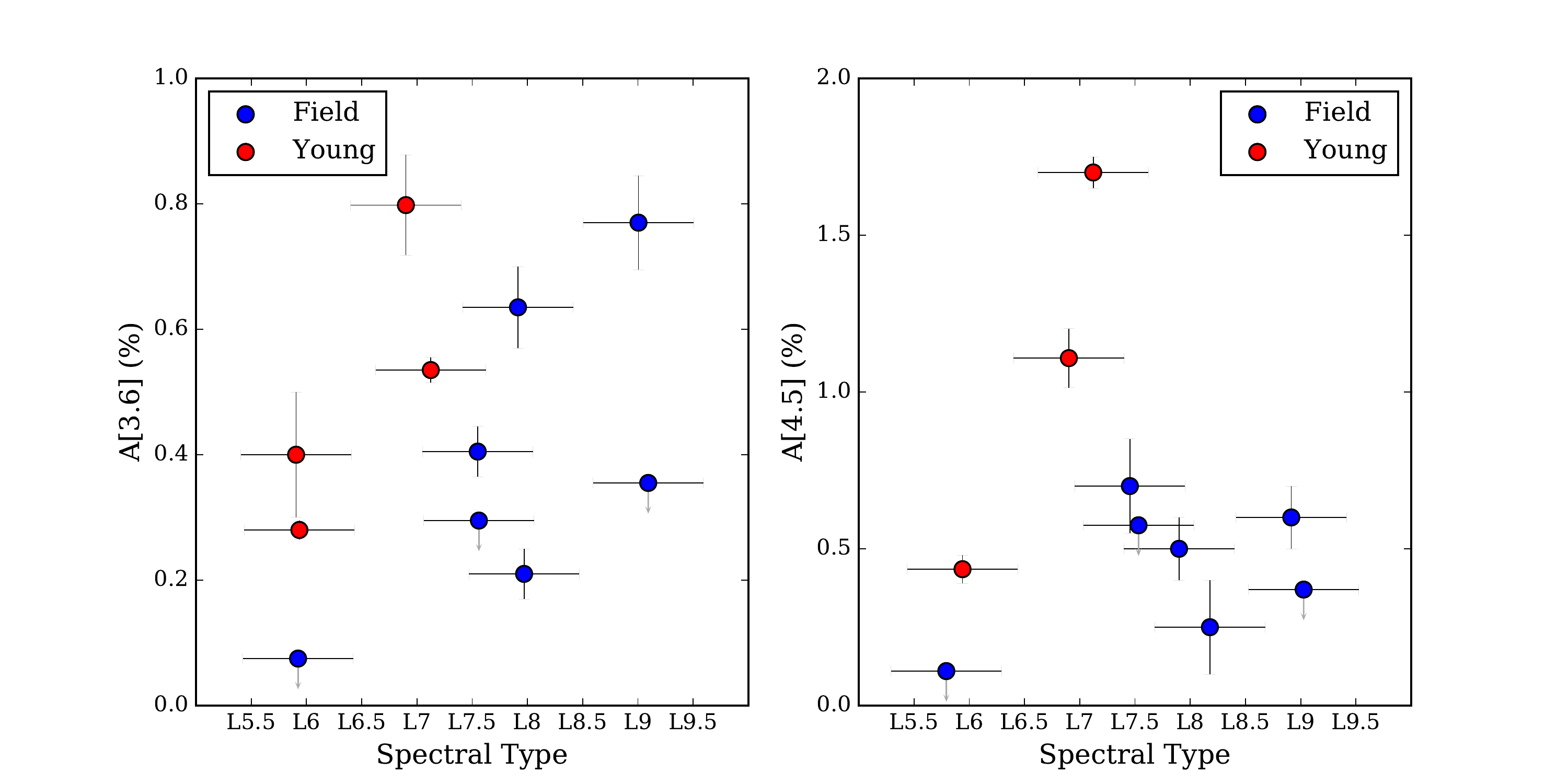}
\caption{{\it Spitzer} [3.6] (left) and [4.5] (right) variability amplitudes versus spectral type for late-type L dwarfs.  Small offsets have been added to the abscissa for differentiation purposes.  2MASS 1119$-$1137AB is not included in this figure.}  
\end{figure*}

\subsection{Brown dwarf rotational evolution}
While the rotation of very young low-mass stars is regulated by interaction with their disks through a magnetic wind,   rotational braking is thought to be extremely inefficient in the substellar regime \citep{bou14}.  However, substellar objects with ages between that of star forming regions and the field have not been well explored.  While a detailed investigation of brown dwarf rotational evolution will require a larger sample of substellar objects at a variety of ages and masses, patterns are already beginning to emerge. The rotation periods of brown dwarfs in the Upper Scorpius association (age = 5-10 Myr; \citealt{herc15}, \citealt{pec16}) measured by \cite{sch15} using K2 span a large range (5-40 hr), with a median of $\sim$1 day, while even younger brown dwarfs typically have rotation periods of several days (e.g., \citealt{joe03}, \citealt{sch04}, \citealt{rod09}, \citealt{cody10}).   These values can be compared to those found for field age and young (10-300 Myr) brown dwarf samples for evidence of rotational evolution.   

To investigate the rotation periods of field age brown dwarfs, we combine the rotation periods from \cite{met15} with the compilations of rotation periods in \cite{cross14} and \cite{vos17}, the rotation period of Luhman 16A (4.5-5.5 hr; \citealt{bue15}) and Luhman 16B (5.05 $\pm$ 0.10 hr; \citealt{burg14}, 4.87 $\pm$ 0.01 hr; \citealt{gill13}) and the two known rotation periods for the Y-type brown dwarfs WISE J140518.39$+$553421.3 (8.54 $\pm$ 0.08 hr; \citealt{cush16}) and WISEP J173835.52$+$273258.9 (6.0 $\pm$ 0.1 hr; \citealt{legg16}). To ensure we do not include low-mass stellar sources in this comparison, we limit our field age sample to those objects with spectral types later than L2 (\citealt{die14}, \citealt{dup17}).  We find that the rotation periods of the 26 brown dwarfs without any signs of low surface gravity range from 1.4 to 11 hours, where 24 of these 26 dwarfs have rotation periods less than 8 hours ($\sim$92\%), with a median rotation period of 4.05 hours.  For younger, low surface gravity brown dwarfs, we consider the low-gravity L dwarfs in Table 2, SIMP J013656.5$+$093347 (2.425 $\pm$ 0.003; \citealt{gagne17b}), recently designated as a member of the $\sim$200 Myr old Carina-Near moving group \citep{gagne17b}, four additional low surface gravity brown dwarfs from \cite{met15} (including 2MASS J13243553$+$6358281, \citealt{gagne18}), and LP261-75B (4.78 $\pm$ 0.95 hr; \citealt{man18}).  We also include in this sample the directly imaged planetary mass companions $\beta$ Pictoris b (8.1 $\pm$ 1.0; \citealt{snell14}, spectral type = L2 $\pm$ 1; \citealt{chilc17}) and 2M1207b (10.7$^{+1.2}_{-0.6}$ hr; \citealt{zhou16}, spectral type = mid-L; \citealt{pat10}).  With a range of rotation periods from 2.4 to 24 hours, we find that only 5 of 14 members of this sample have rotation periods less than 8 hr ($\sim$35\%), with a median rotation period of 9.66 hours.  Thus, brown dwarf rotation periods generally decrease with age, likely because they are spinning up as they contract to their final radii in much the same way young stars do in order to conserve angular momentum.

To determine whether or not the population of brown dwarfs with known rotation periods is consistent with gravitational contraction, we construct a simple model evolutionary track where angular momentum is conserved as a brown dwarf's radius gets smaller with age.  As a starting point, we use the evolutionary models of \cite{bar15} to estimate the radii of a brown dwarfs in Upper Scorpius ($\sim$5 Myr) with masses of 0.01 and 0.08 $M_{\odot}$ and use the maximum and minimum measured periods for this age group.  Assuming angular momentum is conserved, we then calculate rotation periods using theoretical radii from \cite{bar15} for the ages probed in our study.  Figure 6 shows a comparison of all brown dwarf periods for ages $\gtrsim$5 Myr compared to predictions from our simple angular momentum conservation model.  The only brown dwarf that falls outside the range of predicted rotation periods from our model is the young, L4 dwarf 2MASS J16154255$+$4953211, which has a highly uncertain period (see \citealt{met15}).  Otherwise, our model shows general agreement with measured brown dwarf rotation periods, though brown dwarfs with intermediate ages (10-1000 Myr) have not been thoroughly explored.  A larger sample of brown dwarf rotation periods at various ages would help to create a more detailed picture or brown dwarf rotational evolution.    

\begin{figure*}
\plotone{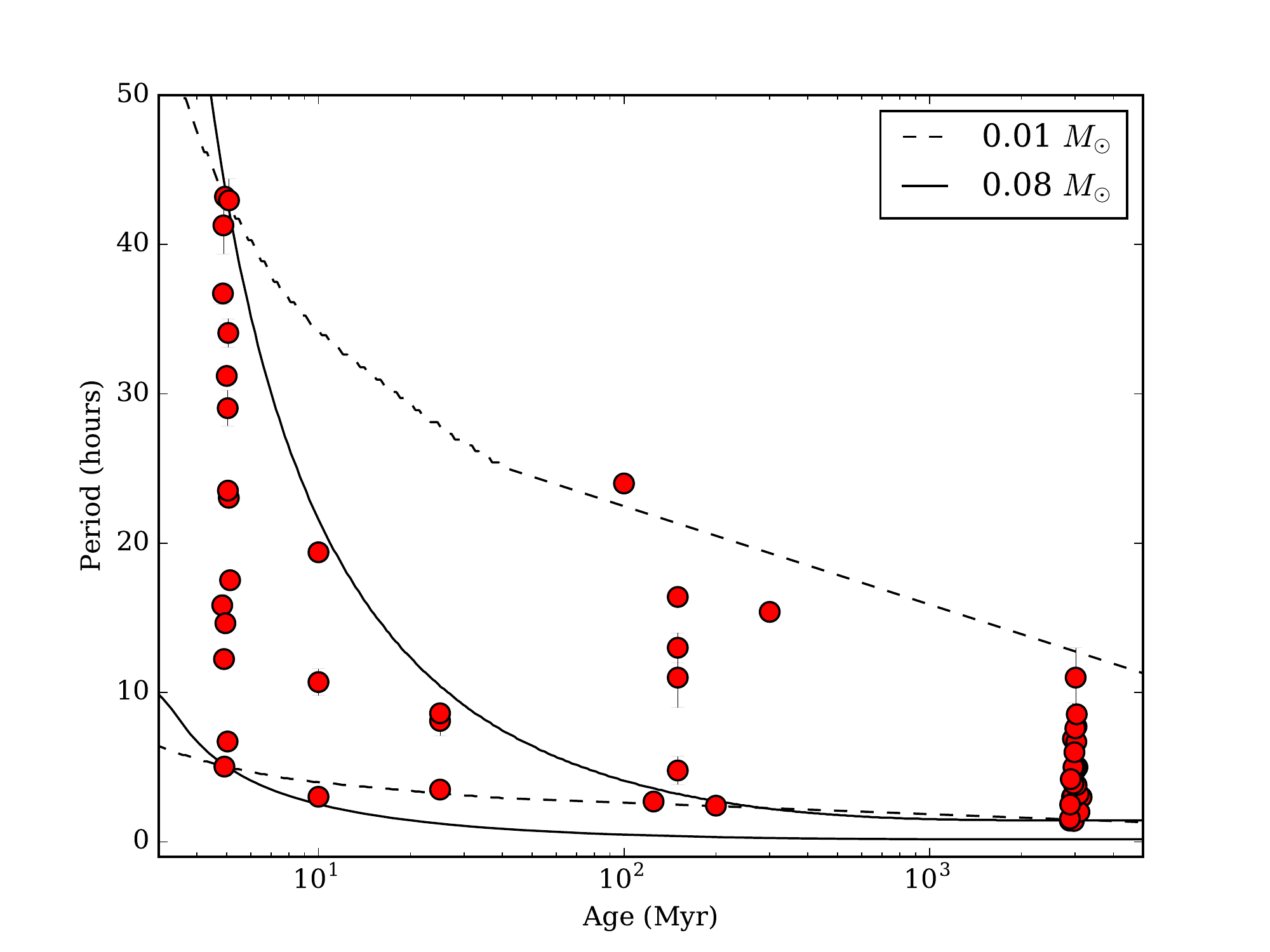}
\caption{The current picture of brown dwarf rotational evolution for ages $\gtrsim$5 Myr.  Field brown dwarfs are assumed to have an age of $\sim$3 Gyr.  The dashed and solid curves represent evolutionary tracks of angular momentum conservation using model radii from \cite{bar15} for masses of 0.01 and 0.08 $M_{\odot}$, respectively.  The initial rotation rate for the upper dashed and solid lines is the maximum period measured for an Upper Scorpius brown dwarf, while the lower lines use the minimum measured period. Small offsets have been added to the abscissa for differentiation purposes.}  
\end{figure*}
 
\section{Conclusion}
We have investigated the photometric variability of the planetary-mass, TW Hya members 2MASS 1119$-$1137AB and WISEA 1147$-$2040 using {\it Spitzer}.  We find a rotation period for WISEA 1147$-$2040 of 19.39$^{+0.33}_{-0.28}$ hours, and find a  period of 3.02$^{+0.04}_{-0.03}$ hours for 2MASS 1119$-$1137AB, which is possibly due to one the rotation of one component of this binary.  We find that WISEA 1147$-$2040 fits into a general trend of longer rotation periods at for brown dwarfs at young ages.  For all other brown dwarfs with measured rotation periods, we find general agreement between measured values and a simple model where brown dwarfs have spun up to conserve angular momentum as they contract with age. 

We have also compared the [3.6] and [4.5] variability amplitudes of young L6-L9 brown dwarfs to field-age brown dwarfs with similar spectral types.  While we find no significant correlation between the amplitude of variability and age as previously seen for L3-L5.5 dwarfs in \cite{met15} at [3.6], we find that young, late-type L dwarfs tend to have higher [4.5] amplitudes than field-age late-type Ls, though with limited confidence ($\sim$75\%). We caution that the sample size used for this comparison is small, and an expanded sample of brown dwarf light curves at different ages would aid in confirming the robustness of this result.     
 
\acknowledgments
We wish to thank the anonymous referee for a helpful report that improved the quality of this work.  A.S.\ and E.S.\ appreciate support from NASA/Habitable Worlds grant NNX16AB62G (PI E. Shkolnik).  This work is based on observations made with the Spitzer Space Telescope, which is operated by the Jet Propulsion Laboratory, California Institute of Technology under a contract with NASA.

\software{{\tt\string photutils} \citep{Bradley2016}, {\tt\string emcee} \citep{fm13}, {\tt\string lifelines} \citep{dav16}}


\begin{thebibliography}{}
\bibitem[Ackerman \& Marley(2001)]{ack01} Ackerman, A.~S., \& Marley, M.~S.\ 2001, \apj, 556, 872 
\bibitem[Allers \& Liu(2013)]{allers13} Allers, K.~N., \& Liu, M.~C.\ 2013, \apj, 772, 79 
\bibitem[Allers et al.(2016)]{allers16} Allers, K.~N., Gallimore, J.~F., Liu, M.~C., \& Dupuy, T.~J.\ 2016, \apj, 819, 133 
\bibitem[Apai et al.(2017)]{apai17} Apai, D., Karalidi, T., Marley, M.~S., et al.\ 2017, Science, 357, 683 
\bibitem[Baraffe et al.(2015)]{bar15} Baraffe, I., Homeier, D., Allard, F., \& Chabrier, G.\ 2015, \aap, 577, A42 
\bibitem[Bell et al.(2015)]{bell15} Bell, C.~P.~M., Mamajek, E.~E., \& Naylor, T.\ 2015, \mnras, 454, 593 
\bibitem[Benneke et al.(2017)]{Benneke2017} Benneke, B., Werner, M., Petigura, E., et al.\ 2017, \apj, 834, 187 
\bibitem[Best et al.(2017)]{best17} Best, W.~M.~J., Liu, M.~C., Dupuy, T.~J., \& Magnier, E.~A.\ 2017, \apjl, 843, L4 
\bibitem[Biller et al.(2018)]{biller18} Biller, B.~A., Vos, J., Buenzli, E., et al.\ 2018, \aj, 155, 95 
\bibitem[Bouvier et al.(2014)]{bou14} Bouvier, J., Matt, S.~P., Mohanty, S., et al.\ 2014, Protostars and Planets VI, 433 
\bibitem[Bowler et al.(2010)]{bowl10} Bowler, B.~P., Liu, M.~C., \& Dupuy, T.~J.\ 2010, \apj, 710, 45 
\bibitem[Bradley et al.(2016)]{Bradley2016} Bradley, L., Sipocz, B., Robitaille, T., et al.\ 2016, Astrophysics Source Code Library, ascl:1609.011 
\bibitem[Buenzli et al.(2014)]{bue14} Buenzli, E., Apai, D., Radigan, J., Reid, I.~N., \& Flateau, D.\ 2014, \apj, 782, 77 
\bibitem[Buenzli et al.(2015)]{bue15} Buenzli, E., Marley, M.~S., Apai, D., et al.\ 2015, \apj, 812, 163 
\bibitem[Burgasser et al.(2002)]{burg02} Burgasser, A.~J., Marley, M.~S., Ackerman, A.~S., et al.\ 2002, \apjl, 571, L151 
\bibitem[Burgasser et al.(2014)]{burg14} Burgasser, A.~J., Gillon, M., Faherty, J.~K., et al.\ 2014, \apj, 785, 48 
\bibitem[Burningham et al.(2010)]{burn10} Burningham, B., Leggett, S.~K., Lucas, P.~W., et al.\ 2010, \mnras, 404, 1952 
\bibitem[Chauvin et al.(2004)]{cha04} Chauvin, G., Lagrange, A.-M., Dumas, C., et al.\ 2004, \aap, 425, L29 
\bibitem[Chauvin et al.(2005)]{cha05} Chauvin, G., Lagrange, A.-M., Dumas, C., et al.\ 2005, \aap, 438, L25 
\bibitem[Chilcote et al.(2017)]{chilc17} Chilcote, J., Pueyo, L., De Rosa, R.~J., et al.\ 2017, \aj, 153, 182 
\bibitem[Chiu et al.(2006)]{chiu06} Chiu, K., Fan, X., Leggett, S.~K., et al.\ 2006, \aj, 131, 2722 
\bibitem[Cody \& Hillenbrand(2010)]{cody10} Cody, A.~M., \& Hillenbrand, L.~A.\ 2010, \apjs, 191, 389 
\bibitem[Crossfield(2014)]{cross14} Crossfield, I.~J.~M.\ 2014, \aap, 566, A130 
\bibitem[Cushing et al.(2016)]{cush16} Cushing, M.~C., Hardegree-Ullman, K.~K., Trucks, J.~L., et al.\ 2016, \apj, 823, 152 
\bibitem[Dahn et al.(2002)]{dahn02} Dahn, C.~C., Harris, H.~C., Vrba, F.~J., et al.\ 2002, \aj, 124, 1170 
\bibitem[Davidson-Pilon(2016)]{dav16} Davidson-Pilon, C.\ 2016, Lifelines, https://github.com/camdavdisonpilon/lifelines 
\bibitem[Deming et al.(2015)]{Deming2015} Deming, D., Knutson, H., Kammer, J., et al.\ 2015, \apj, 805, 132 
\bibitem[Dieterich et al.(2014)]{die14} Dieterich, S.~B., Henry, T.~J., Jao, W.-C., et al.\ 2014, \aj, 147, 94 
\bibitem[Dupuy \& Liu(2017)]{dup17} Dupuy, T.~J., \& Liu, M.~C.\ 2017, \apjs, 231, 15 
\bibitem[Faherty et al.(2013)]{fah13} Faherty, J.~K., Rice, E.~L., Cruz, K.~L., Mamajek, E.~E., \& N{\'u}{\~n}ez, A.\ 2013, \aj, 145, 2 
\bibitem[Faherty et al.(2016)]{fah16} Faherty, J.~K., Riedel, A.~R., Cruz, K.~L., et al.\ 2016, \apjs, 225, 10 
\bibitem[Fazio et al.(2004)]{faz04} Fazio, G.~G., Hora, J.~L., Allen, L.~E., et al.\ 2004, \apjs, 154, 10 
\bibitem[Filippazzo et al.(2015)]{fil15} Filippazzo, J.~C., Rice, E.~L., Faherty, J., et al.\ 2015, \apj, 810, 158 
\bibitem[Foreman-Mackey et al.(2013)]{fm13} Foreman-Mackey, D., Hogg, D.~W., Lang, D., \& Goodman, J.\ 2013, \pasp, 125, 306 
\bibitem[Gagn{\'e} et al.(2014)]{gagne14} Gagn{\'e}, J., Lafreni{\`e}re, D., Doyon, R., Malo, L., \& Artigau, {\'E}.\ 2014, \apj, 783, 121 
\bibitem[Gagn{\'e} et al.(2017a)]{gagne17} Gagn{\'e}, J., Faherty, J.~K., Mamajek, E.~E., et al.\ 2017a, \apjs, 228, 18 
\bibitem[Gagn{\'e} et al.(2017b)]{gagne17b} Gagn{\'e}, J., Faherty, J.~K., Burgasser, A.~J., et al.\ 2017b, \apjl, 841, L1 
\bibitem[Gagn{\'e} et al.(2018)]{gagne18} Gagn{\'e}, J., Allers, K.~N., Theissen, C.~A., et al.\ 2018, arXiv:1802.00493 
\bibitem[Geballe et al.(2002)]{geb02} Geballe, T.~R., Knapp, G.~R., Leggett, S.~K., et al.\ 2002, \apj, 564, 466 
\bibitem[Gillon et al.(2013)]{gill13} Gillon, M., Triaud, A.~H.~M.~J., Jehin, E., et al.\ 2013, \aap, 555, L5 
\bibitem[Gizis et al.(2012)]{giz12} Gizis, J.~E., Faherty, J.~K., Liu, M.~C., et al.\ 2012, \aj, 144, 94 
\bibitem[Gizis et al.(2015)]{giz15} Gizis, J.~E., Allers, K.~N., Liu, M.~C., et al.\ 2015, \apj, 799, 203 
\bibitem[Goodman \& Weare(2010)]{good10} Goodman, J., \& Weare, J.\ 2010, Communications in Applied Mathematics and Computational Science, 5, 65 
\bibitem[Grillmair et al.(2012)]{grill12} Grillmair, C.~J., Carey, S.~J., Stauffer, J.~R., et al.\ 2012, \procspie, 8448, 84481I 
\bibitem[Grillmair et al.(2014)]{grill14} Grillmair, C.~J., Carey, S.~J., Stauffer, J.~R., \& Ingalls, J.~G.\ 2014, \procspie, 9143, 914359 
\bibitem[Herbst et al.(2002)]{herb02} Herbst, W., Bailer-Jones, C.~A.~L., Mundt, R., Meisenheimer, K., \& Wackermann, R.\ 2002, \aap, 396, 513 
\bibitem[Herczeg \& Hillenbrand(2015)]{herc15} Herczeg, G.~J., \& Hillenbrand, L.~A.\ 2015, \apj, 808, 23 
\bibitem[Joergens et al.(2003)]{joe03} Joergens, V., Fern{\'a}ndez, M., Carpenter, J.~M., \& Neuh{\"a}user, R.\ 2003, \apj, 594, 971 
\bibitem[Kaplan \& Meier(1958)]{kap58} Kaplan, E.~L., \& Meier, P.\ 1958, J.\ Am.\ Stat.\ Assoc., 53, 457 
\bibitem[Karalidi et al.(2015)]{kar15} Karalidi, T., Apai, D., Schneider, G., Hanson, J.~R., \& Pasachoff, J.~M.\ 2015, \apj, 814, 65 
\bibitem[Kellogg et al.(2015)]{kell15} Kellogg, K., Metchev, S., Gei{\ss}ler, K., et al.\ 2015, \aj, 150, 182 
\bibitem[Kellogg et al.(2016)]{kell16} Kellogg, K., Metchev, S., Gagn{\'e}, J., \& Faherty, J.\ 2016, \apjl, 821, L15 
\bibitem[Kirkpatrick et al.(1999)]{kirk99} Kirkpatrick, J.~D., Reid, I.~N., Liebert, J., et al.\ 1999, \apj, 519, 802 
\bibitem[Kirkpatrick et al.(2000)]{kirk00} Kirkpatrick, J.~D., Reid, I.~N., Liebert, J., et al.\ 2000, \aj, 120, 447 
\bibitem[Kirkpatrick et al.(2008)]{kirk08} Kirkpatrick, J.~D., Cruz, K.~L., Barman, T.~S., et al.\ 2008, \apj, 689, 1295-1326 
\bibitem[Knutson et al.(2011)]{Knutson2011} Knutson, H.~A., Madhusudhan, N., Cowan, N.~B., et al.\ 2011, \apj, 735, 27 
\bibitem[Leconte(2018)]{lec18} Leconte, J.\ 2018, \apjl, 853, L30 
\bibitem[Leggett et al.(2016)]{legg16} Leggett, S.~K., Cushing, M.~C., Hardegree-Ullman, K.~K., et al.\ 2016, \apj, 830, 141 
\bibitem[Leggett et al.(2017)]{legg17} Leggett, S.~K., Tremblin, P., Esplin, T.~L., Luhman, K.~L., \& Morley, C.~V.\ 2017, \apj, 842, 118 
\bibitem[Liu et al.(2013)]{liu13} Liu, M.~C., Magnier, E.~A., Deacon, N.~R., et al.\ 2013, \apjl, 777, L20 
\bibitem[Lomb(1976)]{lomb76} Lomb, N.~R.\ 1976, \apss, 39, 447 
\bibitem[Looper et al.(2008)]{loop08} Looper, D.~L., Kirkpatrick, J.~D., Cutri, R.~M., et al.\ 2008, \apj, 686, 528-541 
\bibitem[Macintosh et al.(2015)]{mac15} Macintosh, B., Graham, J.~R., Barman, T., et al.\ 2015, Science, 350, 64 
\bibitem[Manjavacas et al.(2018)]{man18} Manjavacas, E., Apai, D., Zhou, Y., et al.\ 2018, \aj, 155, 11 
\bibitem[Marois et al.(2008)]{mar08} Marois, C., Macintosh, B., Barman, T., et al.\ 2008, Science, 322, 1348 
\bibitem[Marois et al.(2010)]{mar10} Marois, C., Zuckerman, B., Konopacky, Q.~M., Macintosh, B., \& Barman, T.\ 2010, \nat, 468, 1080 
\bibitem[Martin et al.(2017)]{mart17} Martin, E.~C., Mace, G.~N., McLean, I.~S., et al.\ 2017, \apj, 838, 73 
\bibitem[Metchev et al.(2015)]{met15} Metchev, S.~A., Heinze, A., Apai, D., et al.\ 2015, \apj, 799, 154 
\bibitem[Patience et al.(2010)]{pat10} Patience, J., King, R.~R., de Rosa, R.~J., \& Marois, C.\ 2010, \aap, 517, A76 
\bibitem[Pecaut \& Mamajek(2016)]{pec16} Pecaut, M.~J., \& Mamajek, E.~E.\ 2016, \mnras, 461, 794 
\bibitem[Radigan et al.(2014)]{rad14} Radigan, J., Lafreni{\`e}re, D., Jayawardhana, R., \& Artigau, E.\ 2014, \apj, 793, 75 
\bibitem[Rodr{\'{\i}}guez-Ledesma et al.(2009)]{rod09} Rodr{\'{\i}}guez-Ledesma, M.~V., Mundt, R., \& Eisl{\"o}ffel, J.\ 2009, \aap, 502, 883 
\bibitem[Scargle(1982)]{scar82} Scargle, J.~D.\ 1982, \apj, 263, 835 
\bibitem[Schmidt et al.(2010)]{schmidt10} Schmidt, S.~J., West, A.~A., Burgasser, A.~J., Bochanski, J.~J., \& Hawley, S.~L.\ 2010, \aj, 139, 1045 
\bibitem[Schneider et al.(2016)]{schneid16} Schneider, A.~C., Windsor, J., Cushing, M.~C., Kirkpatrick, J.~D., \& Wright, E.~L.\ 2016, \apjl, 822, L1 
\bibitem[Schneider et al.(2017)]{schneid17} Schneider, A.~C., Windsor, J., Cushing, M.~C., Kirkpatrick, J.~D., \& Shkolnik, E.~L.\ 2017, \aj, 153, 196 
\bibitem[Scholz \& Eisl{\"o}ffel(2004)]{sch04} Scholz, A., \& Eisl{\"o}ffel, J.\ 2004, \aap, 419, 249 
\bibitem[Scholz(2010)]{sch10} Scholz, R.-D.\ 2010, \aap, 510, L8 
\bibitem[Scholz et al.(2015)]{sch15} Scholz, A., Kostov, V., Jayawardhana, R., \& Mu{\v z}i{\'c}, K.\ 2015, \apjl, 809, L29 
\bibitem[Snellen et al.(2014)]{snell14} Snellen, I.~A.~G., Brandl, B.~R., de Kok, R.~J., et al.\ 2014, \nat, 509, 63 
\bibitem[Tremblin et al.(2016)]{trem16} Tremblin, P., Amundsen, D.~S., Chabrier, G., et al.\ 2016, \apjl, 817, L19 
\bibitem[Vos et al.(2017)]{vos17} Vos, J.~M., Allers, K.~N., \& Biller, B.~A.\ 2017, \apj, 842, 78 
\bibitem[Vos et al.(2018)]{vos18} Vos, J.~M., Allers, K.~N., Biller, B.~A., et al.\ 2018, \mnras, 474, 1041 
\bibitem[Wilson et al.(2014)]{wil14} Wilson, P.~A., Rajan, A., \& Patience, J.\ 2014, \aap, 566, A111 
\bibitem[Zhou et al.(2016)]{zhou16} Zhou, Y., Apai, D., Schneider, G.~H., Marley, M.~S., \& Showman, A.~P.\ 2016, \apj, 818, 176 

\end{thebibliography}
\end{document}